\documentclass[twocolumn,showkeys,floatfix,amsmath,amssymb]{revtex4}
\usepackage{graphicx}% Include figure files
\usepackage{dcolumn}% Align table columns on decimal point
\usepackage{color}
\usepackage{bm}% bold math
\usepackage{tabularx}
\usepackage{physics}
\usepackage{amsmath}
\usepackage{mathastext} %No tilting for maths

\begin{document}

%\title{Engineering Giant Perpendicular Magnetocrystalline Anisotropy at Fe/Pb(001) interface}
\title{Large Perpendicular Magnetocrystalline Anisotropy at Fe/Pb(001) interface}

\author{Xiaoxuan Ma and Jun Hu}
\email[]{E-mail: jhu@suda.edu.cn}
\affiliation{College of Physics, Optoelectronics and Energy, Soochow University, Suzhou, Jiangsu 215006, China \\
Jiangsu Key Laboratory of Thin Films, Soochow University, Suzhou, Jiangsu 215006, China.}

\begin{abstract}
Search for ultrathin magnetic film with large perpendicular magnetocrystalline anisotropy (PMA) has been inspired for years by the continuous miniaturization of magnetic units in spintronics devices. The common magnetic materials used in research and applications are based on Fe because the pure Fe metal is the best yet simple magnetic material from nature. Through systematic first-principles calculations, we explored the possibility to produce large PMA with ultrathin Fe on non-noble and non-magnetic Pb(001) substrate. Interestingly, huge magnetocrystalline anisotropy energy (MAE) of 7.6 meV was found in Pb/Fe/Pb(001) sandwich structure with only half monolayer Fe. Analysis of electronic structures reveals that the magnetic proximity effect at the interface is responsible for this significant enhancement of MAE. The MAE further increases to 13.6 meV with triply repeated capping Pb and intermediate Fe layers. Furthermore, the MAE can be tuned conveniently by charge injection.%Therefore, this structure is of great potential in applications in spintronics devices at high temperature.
\end{abstract}

\maketitle

\section{Introduction}
Magnetic tunnel junctions (MTJs) are the key building blocks in spintronics devices for the applications in modern technologies such as data storage and spin valve. \cite{Science2001,RMP2004,HeinrichAJ,LiuLQ,HeinrichBW,IEEETM2010} A typical MTJ consists of two ferromagnetic Fe thin films separated by ultrathin MgO intermediate layer. \cite{HeinrichB-1,HeinrichB-2,ButlerWH} As the spintronics devices keep miniaturizing rapidly in the recent decade, MTJs with sizes scaling down to a few nanometers are desired.  \cite{HeinrichAJ,IEEETM2010} To this end, the thickness of the ferromagnetic thin films in MTJs down to a few atomic layers is preferred. In this realm, large perpendicular magnetocrystalline anisotropy (PMA) is the most critical prerequisite, because it overcomes the thermal fluctuation of spins and in-plane shape magnetic anisotropy to maintain perpendicular magnetization at high temperature. \cite{RMP2017}

Generally, the magnetocrystalline anisotropy (MA) derives from the spin-orbit coupling (SOC). However, the strong magnetic materials such as 3d transition metals (TMs) which are commonly used in the MTJs usually display weak SOC. Therefore, heavy rare-earth and noble TM elements (such as Gd, Tb, Pd and Pt) were alloyed with 3d TMs (especially Fe and Co), in order to enhance the SOC through the interaction between the light and heavy TM atoms. \cite{RE-TM-1,RE-TM-2,L10-FePt,L10-CoPt} Interestingly, large PMA were successfully achieved in a series of such alloys as expected. For example, the magnetocrystalline anisotropy energies (MAEs) in L1$_0$ FePt and CoPt crystals are respectively enhanced to 2.7 and 1.0 meV per formula unit, hundreds of times of the intrinsic MAEs in bulk Fe and Co ($\sim10^{-3}$ meV per atom). \cite{L10-MAE-1,L10-MAE-2,L10-MAE-3} The MAEs are even larger in ultrathin films, due to the more localized surface states and quantum confinement. It was found that the MAE can be as large as $~5.2$ meV/Fe atom in the sandwich structure Pt/Fe/Pt(001) where the thicknesses of both the capping Pt layer and ferromagnetic Fe layer are only one monolayer (ML). \cite{Pt/Fe/Pt-1,Pt/Fe/Pt-2} However, less expensive but more abundant elements such as Mo and W are more attractive for real applications. Unfortunately, the Fe/Mo(110) and Fe/W(110) systems show only in-plane MA. \cite{Fe/W,Fe/Mo} Parallelly, alloys and multilayers with only 3d TM elements were also investigated, but their MAEs are much smaller than that of Pt/Fe/Pt(001). \cite{CoFeB-1,CoFeB-2, CoFeB-3, CoFeB-4, CoFeB-5, Freeman1, Ito2015} Therefore, it is of great interest to search and design new magnetic materials without noble elements but with large PMA.

Besides TM elements, the heavy {\it p}--valent elements such as Pb possess strong SOC. The SOC constant of Pb--6{\it p} orbitals is $\sim0.9$ eV \cite{SOC-Pb}, much larger than that of Pt--5{\it d} orbitals ($\sim0.6$ eV) \cite{SOC-Pt}.  Intuitively, the SOC effect in Fe may be enhanced significantly through alloying Fe with Pb or building Fe/Pb multilayers. Note that bulk Pb crystallizes face-centered cubic (FCC) phase with lattice constant of 4.95 \AA, so its (001) surface has favorable lattice constant ({\it a} = 3.50 \AA) to match the $\sqrt{2}\times\sqrt{2}$ $\gamma$--Fe(001) surface ({\it a$^\prime$} = 3.59 {\AA}). In this paper, we investigated the magnetic properties of Fe/Pb(001) multilayers, based on first-principles calculations. Interestingly, alternately placing two MLs Pb and half ML Fe by three times on Pb(001) possesses huge MAE of 13.6 meV, and the MAE can be further enhanced to 16.5 meV by injecting hole charge. Analysis of electronic properties reveals that the magnetic proximity effect at the Fe/Pb interface plays an important role in the enhancement of the MAE.

\section{Computational details}
We firstly studied the Fe/Pb(001) bilayer system, with six-layer Pb slab to model the Pb(001) substrate and a few ML(s) $\sqrt{2}\times\sqrt{2}$ $\gamma$--Fe(001) on it. The combined bilayer system is denoted as $Fe_n/Pb$, where {\it n} is the number of Fe atoms in the unit cell of the slab. Density functional theory (DFT) calculations were carried out with the Vienna {\it ab-initio} simulation package. \cite{VASP1,VASP2} The interaction between valence electrons and ionic cores was described within the framework of the projector augmented wave (PAW) method. \cite{PAW1,PAW2} The spin-polarized generalized-gradient approximation (GGA) was used for the exchange-correlation potentials. \cite{PBE} The energy cutoff for the plane wave basis expansion was set to 350 eV. For all slabs  a vacuum of about 15 {\AA} along the surface normal was inserted between neighboring slabs to mimic the two-dimensional periodicity. The two-dimensional Brillouin zone was sampled by a $45\times45$ k-grid mesh. The atomic positions except the bottom two layers were fully relaxed using the conjugated gradient method until the force on each atom is smaller than 0.01 eV/{\AA}.  We have checked the results using 13-layer Pb slab, which shows that six-layer Pb slab is good enough in this work.

\section{Results and discussions}

\begin{figure}
\centering
\includegraphics[width=8.5cm]{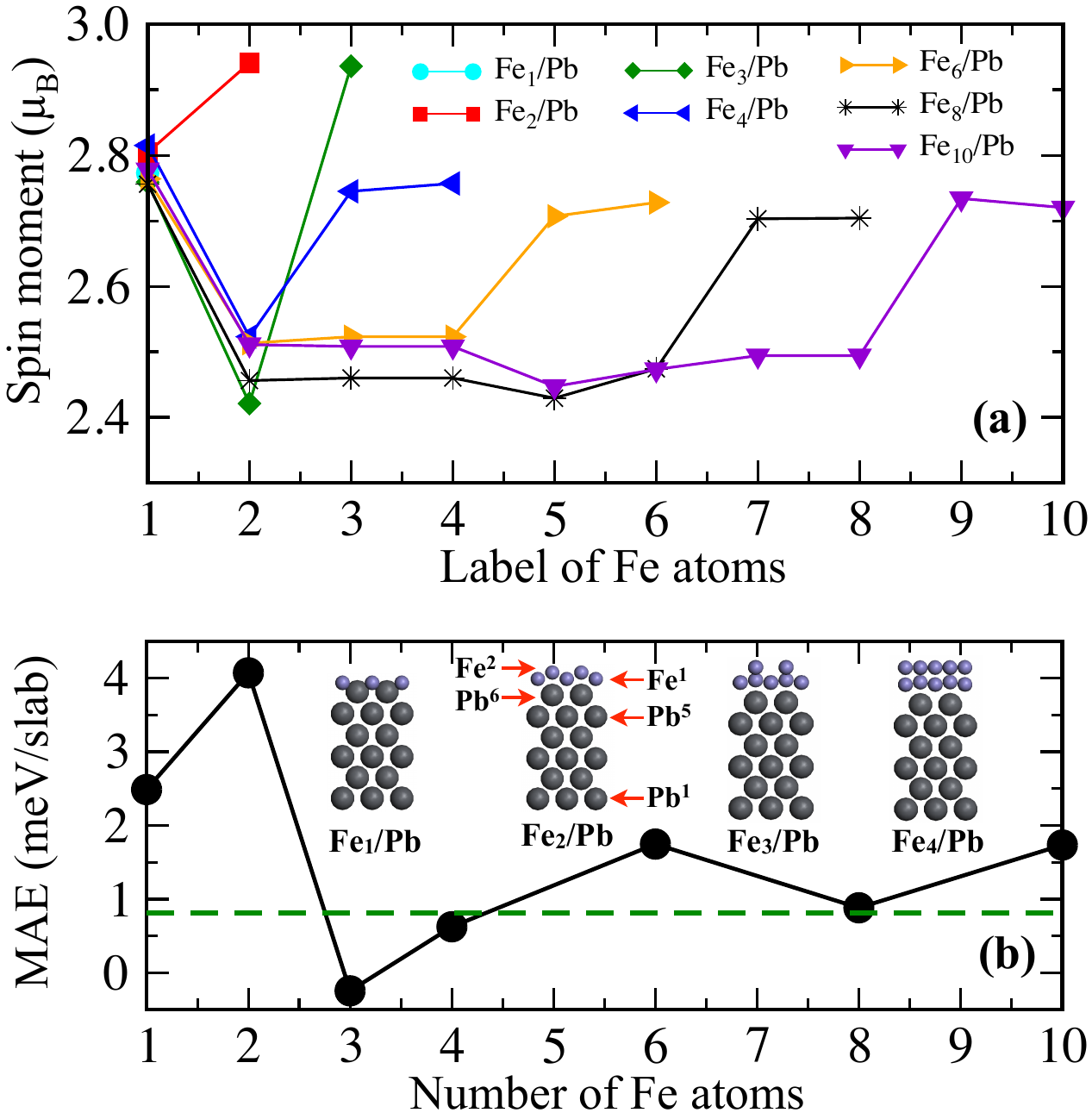}
\caption{(Color online) Magnetic properties of $Fe_n/Pb$ slabs. (a) Local spin moments on the Fe atoms. (b) Magnetocrystalline anisotropy energy (MAE) of each slab. The green dashed line indicates the MAE of Fe monolayer (0.8 meV) as a benchmark. The insets show the relaxed structures for $n=1\sim4$. The grey and purple spheres stand for Pb and Fe atoms, respectively.
}\label{mae1}
\end{figure}

After relaxation of atomic positions, the interfacial Fe layer is significantly buckled, due to the large difference of atomic sizes between Fe and Pb atoms. The Fe-Pb bond length between the closest Fe and Pb atoms at the interface is around 2.75 {\AA}, manifesting strong interaction between the Fe and Pb atoms. For example, the binding energy of $Fe_1/Pb$ is 1.7 eV/Fe. All the Fe atoms are spin polarized. As shown in Fig. \ref{mae1}(a), the local spin moment on an Fe atom ($M_{S,Fe}$) depends on its location and can be sorted into three groups. The outermost Fe atoms in $Fe_2/Pb$ and $Fe_3/Pb$ have the largest $M_{S,Fe}$ of 2.94 $\mu_B$, while the outermost Fe atoms in other slabs with ${\it n}\geqslant4$ are $2.73\pm0.03~\mu_B$. In addition, the $M_{S,Fe}$ on the interfacial Fe atom [denoted as $Fe^1$ in Fig. \ref{mae1}(b)] which is also the outermost Fe atom in $Fe_1/Pb$ are almost the same for all slabs ($2.79\pm0.03~\mu_B$). Finally, the Fe atoms of the intermediate layers possess smaller $M_{S,Fe}$ of $2.47\pm0.05~\mu_B$. This situation is similar to the pure Fe thin film where the $M_{S,Fe}$ of the inner and surface Fe atoms are 2.25 and 2.95 $\mu_B$, respectively. On the other hand, spin polarization is induced on the interfacial Pb atom [$Pb^6$ as indicated in Fig. \ref{mae1}(b)], which results in the local spin moment of $-0.07~\mu_B$. Nevertheless, the magnetic proximity effect in $Fe_n/Pb$ is much weaker than that in $L1_0$ FePt alloy where the induced spin moment on the Pt atom is as large as $0.35~\mu_B$. \cite{L10-MAE-1}

Then we calculated the MAEs of the $Fe_n/Pb$ slabs. Conventionally, the MAEs is evaluated by the change of the total energy for the magnetization switching between orientations parallel and perpendicular to the surface plane \cite{Pt/Fe/Pt-1,Pt/Fe/Pt-2}, $MAE=E_\parallel-E_\perp$, where $E_\parallel$ and $E_\perp$ denote the corresponding total energies, respectively. This method require extremely fine sampling meshes for the k-space integrations, which leads to bad convergence of MAE to the number of k points. On the contrary, the torque method proposed by Wang {\it et~al.} \cite{Torque1} is much less sensitive to the number of k points. Instead of directly calculating the total energy difference, the torque method integrates the torque of the SOC Hamiltonian ($H_{SO}$)
\begin{equation}
MAE=\sum_{n{\bf k}}^{occ} \ev**{\pdv{H_{SO}}{\theta}}{\Psi_{n{\bf k}}}_{\theta=45^{\circ}}.
\end{equation}
Here, $\Psi_{n{\bf k}}$ is the {\it n}th relativistic eigenvector at {\bf k} point, and $\theta$ is the angle between the spin orientation and surface normal. In the spin space, $H_{SO}$ can be expressed as a $2\times2$ matrix. Therefore, we can extract the contributions from the diagonal and off-diagonal elements of $H_{SO}$ matrix and divide the MAE into three parts: MAE(uu), MAE(dd) and MAE(ud$+$du). Here, `uu', `dd' and `ud+du' represent the contributions from the coupling between majority spin states, minority spin states, and cross-spin states, respectively. Recently, we implemented this method in the framework of PAW and showed that the torque method is quite efficient for calculating MAEs of magnetic materials. \cite{Torque2,Hu-NL-2014} From Fig. \ref{mae1}(b) it can be seen that the MAEs of $Fe_1/Pb$ and $Fe_2/Pb$ are significantly enhanced to 2.5 and 4.1 meV, respectively, compared to the pure Fe thin film (0.8 meV). \cite{MAE-Fe} In contrast, the MAE of $Fe_3/Pb$ becomes -0.3 meV, implying in-plane magnetocrystalline anisotropy. The MAEs of the other slabs vary between 0.65 and 1.75 meV. Clearly, the different MAEs are associated with the different structures of the Fe thin films. For $Fe_n/Pb$ thin films with ${\it n}\geqslant4$, the interaction between the Fe atoms dominate their magnetic properties, hence the MAEs oscillate near the intrinsic value of pure Fe thin films. On the contrary, the interactions between the interfacial Fe and Pb atoms are dominant in the MAEs for $Fe_1/Pb$ and $Fe_2/Pb$.

\begin{figure}
\centering
\includegraphics[width=8.5cm]{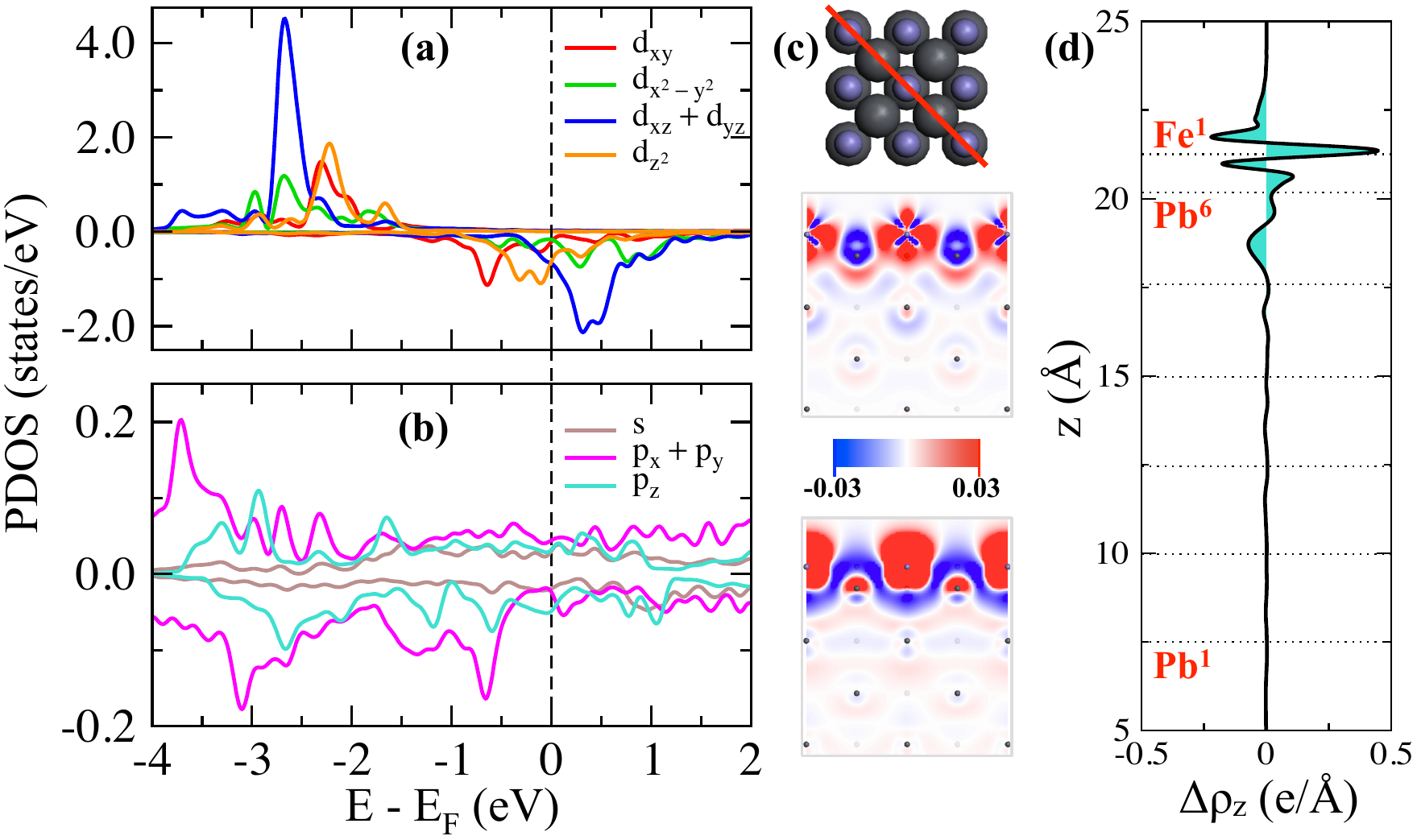}
\caption{(Color online) Electronic properties of $Fe_1/Pb$. (a) and (b) Projected density of states (PDOS) of $Fe^1$ and $Pb^6$, respectively. The positive and negative signs stand for majority and minority spins, respectively. (c) Top panel: top view of the atomic structure; middle panel: electron charge redistribution caused by the interaction between the Fe and Pb atoms [$\Delta\rho=\rho(Fe_1/Pb)-\rho(Pb)-\rho(Fe)$]; bottom panel: spin density ($\rho_s=\rho_{\uparrow}-\rho_{\downarrow}$). Both $\Delta\rho$ and $\rho_s$ are on (110) plane indicated by the red line in top panel, with cutoff of $\pm~0.03~e/{\AA}^3$. The grey and purple spheres stand for Pb and Fe atoms, respectively. (d) Planar summation of electron charge redistribution ($\Delta\rho_z$) along the surface normal ({\it z}).
}\label{dos1}
\end{figure}

To illustrate the effect of the interfacial interactions on the MAE, we calculated the projected density of states (PDOS) of $Fe_1/Pb$ as shown in Fig. \ref{dos1}. Here the Fe atom locates above the hollow site composed of four Pb atoms with Fe--Pb bond length of 2.70 {\AA}, which imposes local symmetry of $C_{4v}$ around the Fe atom. Therefore, the $Fe-3d$ orbitals split into four group: $d_{xy}$, $d_{x^2-y^2}$, $d_{xz/yz}$ and $d_{z^2}$, with the $d_{xz}$ and $d_{yz}$ keeping degenerate. From Fig. \ref{dos1}(a), it can be seen that the $d_{xz/yz}$ orbitals have sharp and localized PDOS, which indicates that they do not interact with Pb atoms notably. In contrast, the other $Fe-3d$ orbitals are more delocalized, because they interact with the broadly dispersive $6p$ orbitals of $Pb^6$ as seen in Fig.\ref{dos1}(b). Moreover, the electron charge redistribution in Fig. \ref{dos1}(c) and \ref{dos1}(d) indicates that the Fe atoms gain electron charge from $Pb^6$ and the electron charge accumulates at the Fe layer.  Integrating the PDOS of each orbital of the Fe atom up to the Fermi energy ($E_F$) yields the electron occupancies of 1.6{\it e}, 1.2{\it e}, 2.2{\it e} and 1.5{\it e} on the $d_{xy}$, $d_{x^2-y^2}$, $d_{xz/yz}$ and $d_{z^2}$ orbitals, respectively. Therefore, the total electron charge on the $Fe-3d$ orbitals is 6.5{\it e}, which indicates the accumulated electron charge on each Fe atom is 0.5{\it e}. Furthermore, the $d_{xz/yz}$ orbital is nearly completely spin-polarized with local spin moment of $\sim1.7~\mu_B$, contributing about 60\% to the total spin moment. The local spin moments contributed by the other Fe--3d orbitals are much smaller, about 0.4, 0.7 and 0.4 $\mu_B$ by $d_{xy}$, $d_{x^2-y^2}$ and $d_{z^2}$, respectively. Similarly, in $Fe_2/Pb$ the Fe$-3d$ and Pb$-6p$ orbitals of hybridize with each other strong, as indicated by the PDOS in Fig. S1 in the Supporting Information.

Apparently, the interfacial Pb atom plays an important role in the enhancement of the MAEs of $Fe_1/Pb$ and $Fe_2/Pb$, even though the induced spin moments on the Pb atom are small. Therefore, it is instructive to estimate the contributions of individual  atoms quantitatively. For this purpose, we rewrite Eq. (1) by inserting atomic orbitals
\begin{equation}
MAE=\sum_{n{\bf k}}^{occ} \sum_i \abs{\it C_i}^2 \ev**{\pdv{H_{SO}}{\theta}}{\varphi_i}_{\theta=45^{\circ}},
\end{equation}
where ${\it C_i}=\bra{\varphi_i}\ket{\Psi_{n{\bf k}}}$ is the projected coefficient of an atomic orbital ($\varphi_i$) on the eigenvector of the system ($\Psi_{n{\bf k}}$). Summing over all the atomic orbitals of a selected atom yields the atomic resolved MAE.
%General equation should be expressed as follows, but $\bra{\Psi_{n{\bf k}}}\ket{\varphi_i} \bra{\varphi_j}\ket{\Psi_{n{\bf k}}}$ reduces to $\abs{\it C_i}^2$, because $H_{SO}$ localizes on each atom.
%\begin{equation}
%MAE=\sum_{n{\bf k}}^{occ} \sum_{i,j} \bra{\Psi_{n{\bf k}}}\ket{\varphi_i} \bra{\varphi_j}\ket{\Psi_{n{\bf k}}} \ev**{\pdv{H_{SO}}{\theta}}{\varphi_i}_{\theta=45^{\circ}}
%\end{equation} 

\begin{figure}
\centering
\includegraphics[width=8.5cm]{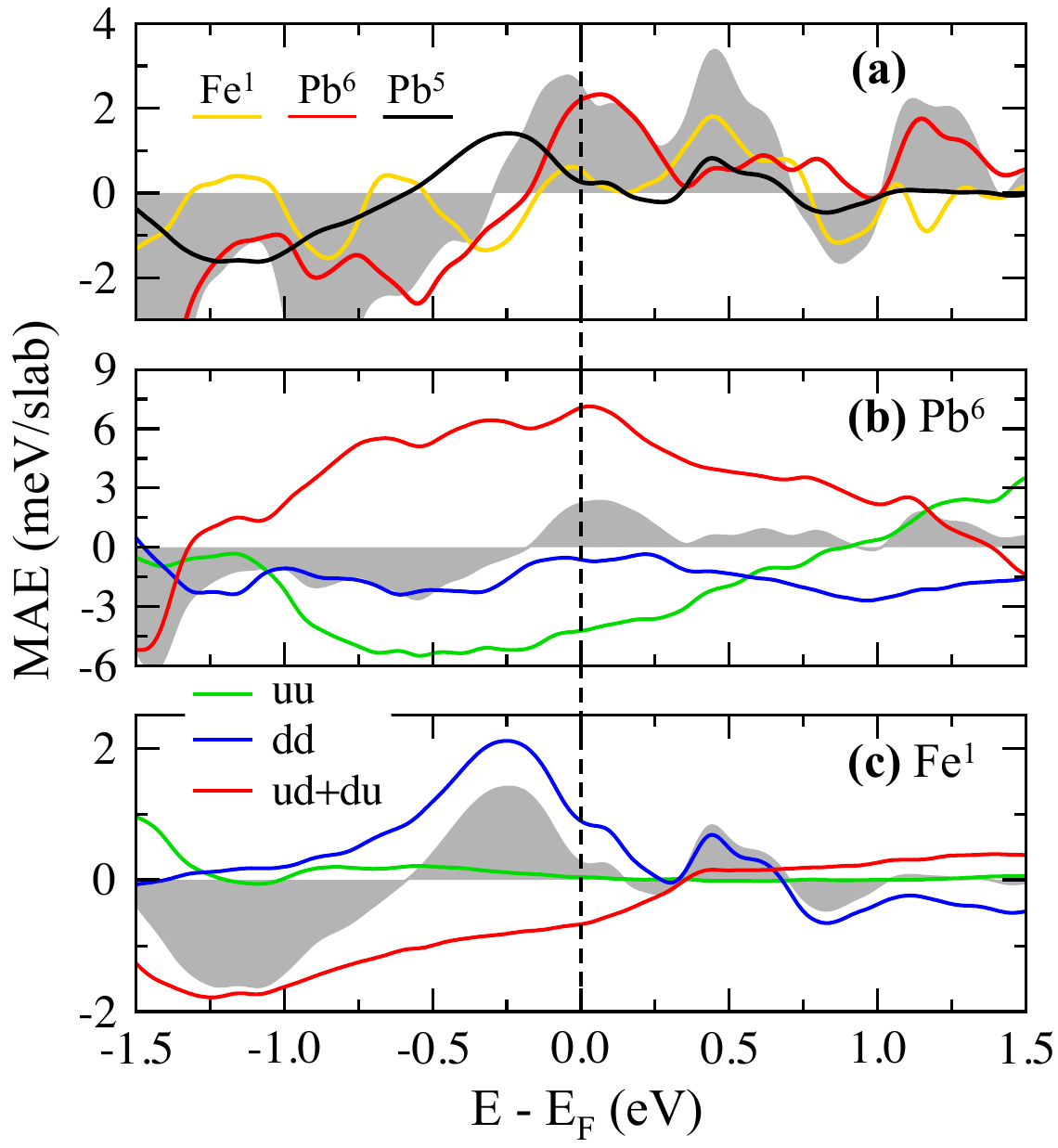}
\caption{(Color online) Atom resolved MAEs in $Fe_1/Pb$. (a) The total (grey area)  MAE of the slab and local MAEs on the interfacial atoms. (b) and (c) The spin resolved MAEs of $Pb^6$ and $Fe^1$. Here the grey area stands for the total local MAE. `uu', `dd' and `ud+du' notate the contributions from the coupling between majority spin states, minority spin states, and cross spin states, respectively.
}\label{torque}
\end{figure}

Interestingly, we found that the interfacial Pb atoms have much larger contribution to the total MAE than the Fe atom. For $Fe_1/Pb$, $Pb^6$ contributes MAE of 2.20 meV which is 88\% of the total MAE [see Fig. \ref{torque}(a)], while the MAE contributed by $Fe^1$ is only 0.26 meV, and the MAEs from other Pb atoms are negligible. For $Fe_2/Pb$, it can be seen from Fig. S1 in the Supporting Information that, both $Pb^6$ and $Pb^5$ (interfacial Fe and Pb atoms) have large contributions to MAE, 1.75 and 1.44 meV, respectively, accounting for 78\% of the total MAE. The outer Fe atom ($Fe^2$) also contributes to MAE largely (0.98 meV), but the MAE from $Fe^1$ is negligible (0.08 meV). This is different from Pt/Fe/Pt(001) where the Fe layer has dominant contribution (5.50 meV/Fe) to the total MAE (5.21 meV/slab), while the Pt layer has negative and smaller contribution ($-0.85$ meV). \cite{Pt/Fe/Pt-1}

To get a deeper insight into the large MAEs from the interfacial Pb atoms, we extracted the spin-resolved MAEs and plotted the $E_F$-dependent curves in Fig. \ref{torque}(b). It can be seen that the total MAE near the $E_F$ mainly originates from the competition between negative MAE(uu) and positive MAE(ud$+$du). However, it is difficult to further distinguish the contributions of $Pb-6p$ orbitals to MAE(uu) and MAE(ud$+$du), because their energy bands quite dispersive as shown in Fig. S2 in the Supporting Information. As for $Fe^1$, Fig. \ref{torque}(c) shows that the MAE comes from the competition between negative MAE(dd) and positive MAE(ud$+$du), while the MAE(uu) contributes little because the majority spin states are fully occupied [Fig. \ref{dos1}(a)]. Clearly, the negative MAE(dd) and positive MAE(ud$+$du) near the $E_F$ have comparable amplitudes, so they cancel each other and result in small contribution to the total MAE.

\begin{figure}
\centering
\includegraphics[width=8.5cm]{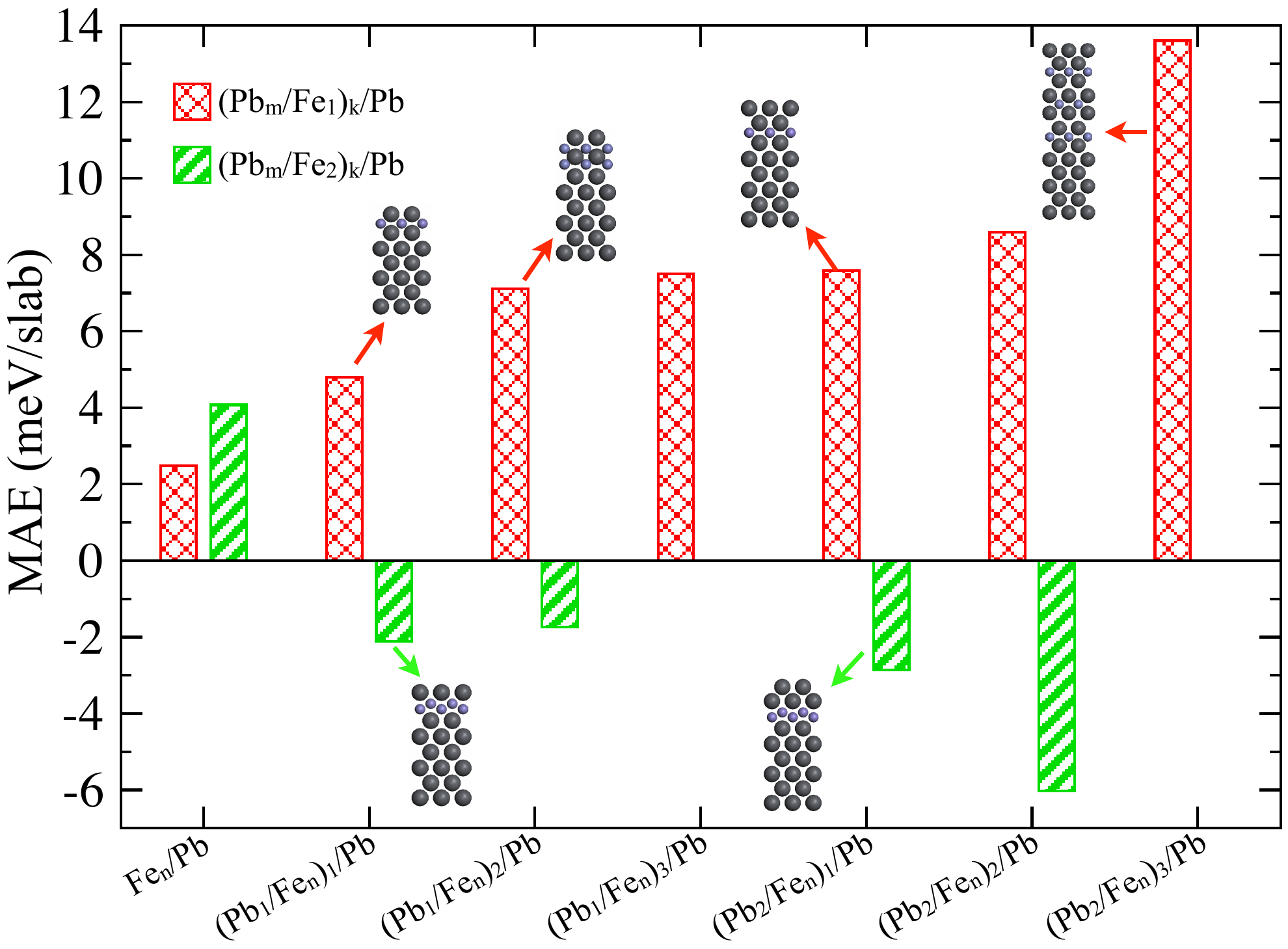}
\caption{(Color online) MAEs of slabs $(Pb_m/Fe_n)_k/Pb$. The MAEs of $Fe_1/Pb$ and $Fe_2/Pb$ are plotted for reference. The insets show the atomic structures.
}\label{mae2}
\end{figure}

This unexpected result indicates that the magnetic proximity effect which induces small amount of spin polarization on the interfacial Pb atoms is the essential factor for the significant enhancement of the MAEs in $Fe_1/Pb$ and $Fe_2/Pb$. Therefore, large MAE is expected if Pb layers exist on both sides of the Fe layer. We considered a series of slab models with repeating structure of Pb capping layer on the Fe layer, denoted as $(Pb_m/Fe_n)_k/Pb$. Here, {\it m} and {\it n} are the thicknesses of the capping Pb layer and Fe layer, and {\it k} denotes the repeating times. From Fig. \ref{mae2}, it can be seen that capping extra Pb layer on $Fe_2/Pb$ turns the large positive MAE into negative, implying that the $(Pb_m/Fe_2/)_k/Pb$ prefers in-plane magnetization. On the contrary, the capping Pb layer boosts the MAE dramatically for $(Pb_m/Fe_1/)_k/Pb$. Capping 1 ML Pb on $Fe_1/Pb$ makes the MAE increase from 2.49 meV to 4.79 meV. Repeating the $Pb_1/Fe_1$ bilayer by 2 and 3 times further augment the MAEs to 7.11 and 7.50 meV, respectively. More dramastically, capping 2 MLs Pb on $Fe_1/Pb$ (i.e. $Pb_2/Fe_1/Pb$) results in MAE of 7.59 meV, which is greatly larger than that of $Pt/Fe/Pt$ (5.21 meV). The atom-resolved MAEs in Fig. S3 in Supporting Information shows that the two Pb layers adjacent to the Fe layer have comparable and dominant contributions (totally 5.75 meV) to the total MAE. Repeating the $Pb_2/Fe_1$ trilayer by 2 and 3 times further enhances the MAEs to 8.59 and 13.60 meV, respectively. Here, we took the sandwich structure $Pb_2/Fe_1/Pb$ to test the convergence of MAE to the number of k points. By calculating the MAE with different k-grid meshes through both the torque method and total energy differences, we found that the MAE from the torque method converges around k-grid mesh of 40$\times$40, but that from the total energy difference is still far from convergence up to 45$\times$45, as seen in Fig. S4 in the Supporting Information. This bad convergence behavior is mainly caused by the situation that there are many bands crossing the Fermi level, as shown in Fig. S5 in Supporting Information. In fact, the convergence problem is a long-term issue, especially for the calculations of properties related to the SOC effect. \cite{Yao-bcc-Fe}

%However, the MAEs per $Pb_2/Fe_1$ trilayer are 4.30 and 4.37 meV in the later two cases, notably smaller than that in $Pb_2/Fe_1/Pb$.

\begin{figure}
\centering
\includegraphics[width=8.5cm]{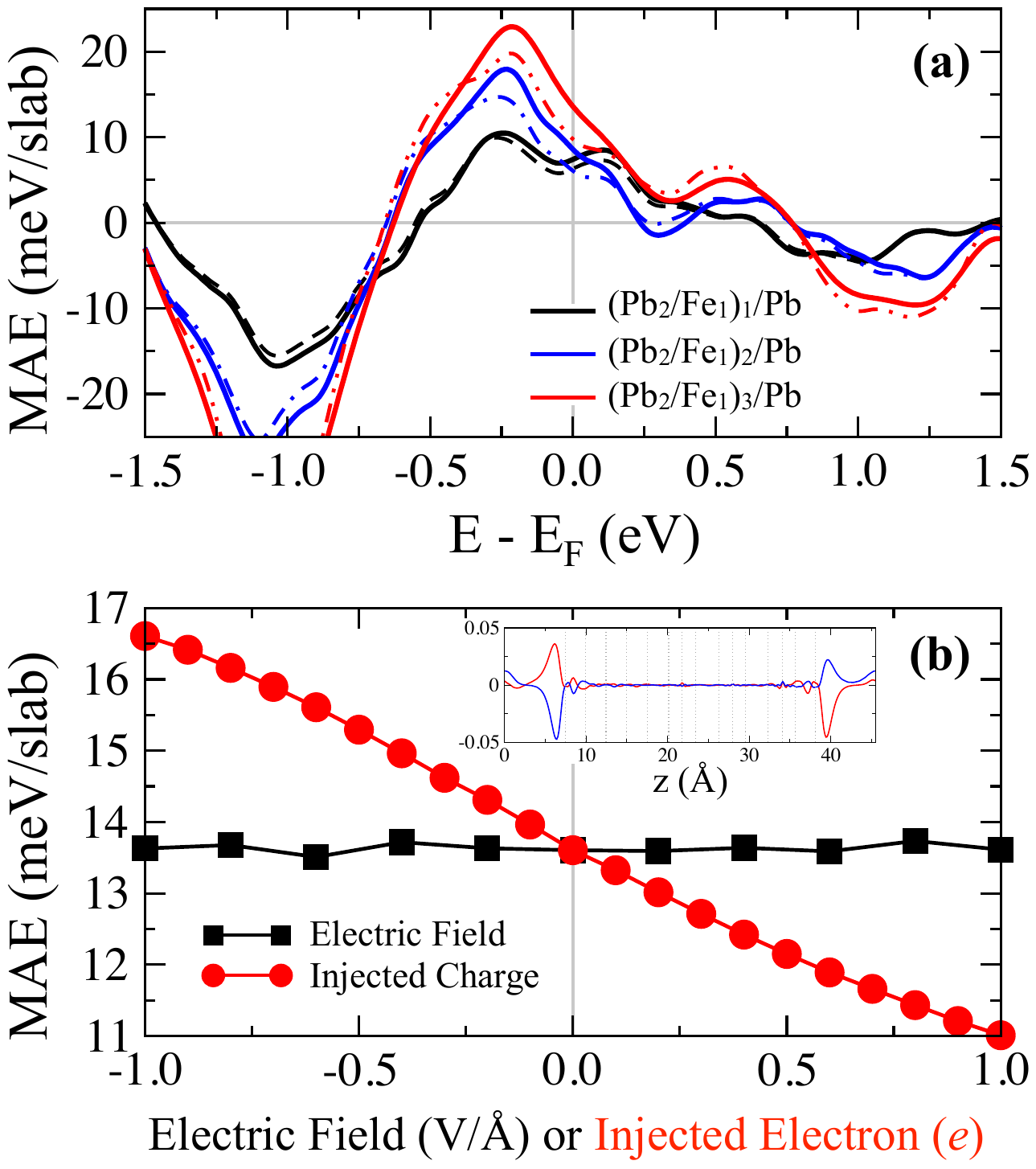}
\caption{(Color online) (a) The $E_F-$dependent MAEs of $(Pb_2/Fe_1)_k/Pb$ with $k=1\sim3$. The dashed curves are the MAE contributed by the Pb atoms adjacent to the Fe layer of each slab. (b) MAEs of $(Pb_2/Fe_1)_3/Pb$ under electric field (diamond) and injected electrons (circles). Positive electric field goes anti-parallelly to the surface normal of the slabs. The inset shows the electron charge redistribution along the surface normal ({\it z}) caused by the positive (red curve) and negative (blue curve) electric fields [$\Delta\rho=\rho(\varepsilon=\pm1~V/\AA)-\rho(\varepsilon=0)$].
}\label{torque2}
\end{figure}
	
Interestingly, from Fig. \ref{torque2}(a), we can see that the Pb atoms adjacent to the Fe layer contribute most part of the MAE for $(Pb_2/Fe_1/)_k/Pb$, similar with the situation in $Fe_1/Pb$. Furthermore, we can see that the MAEs change rapidly near the $E_F$ for $(Pb_2/Fe_1/)_2/Pb$ and $(Pb_2/Fe_1/)_3/Pb$, which is associated with the significant change of the electron occupancies on the orbitals near the $E_F$. \cite{Torque2,Tsymbal} Therefore, if the $E_F$ can be controlled to shift upwards or downwards, then the MAEs will be reduced or enhanced. Usually, this can be achieved by applying external electric field or injecting charge. \cite{Pt/Fe/Pt-1,Pt/Fe/Pt-2,Tsymbal} In our calculations, an external electric field is introduced by adding the corresponding electrostatic potential to the slab with planar dipole layer method \cite{E-field} and the dipolar correction is included. An electric field is assumed to be positive when it is antiparallel to the surface normal. The charge doping is controlled by the number of valence electrons and a uniformly charged background is used to neutralize the periodic slab. Typically, by adding (removing) small part of the valence electrons such as 1 $e$ per unit cell, one can simulate a negative (positive) charge doping.

Unfortunately, the electric field has little influence on the MAE as seen in Fig. \ref{torque2}(b). This is because the screening effect strongly suppresses the penetration of the electric field into the slab, so that only the outermost charge is disturbed [see the inset in Fig. \ref{torque2}(b)]. Accordingly, the electron occupancies do not change visibly under external electric field. In contrast, charge injection through gate voltage is not restricted by the screening effect, so it has much more remarkable impact on the MAEs of $(Pb_2/Fe_1/)_2/Pb$ and $(Pb_2/Fe_1/)_3/Pb$, as displayed in Fig. \ref{torque2}(b). It can be seen that injecting holes into (or extracting electrons from) the slab makes the MAE of $(Pb_2/Fe_1/)_3/Pb$ increase significantly and the MAE reaches 16.5 meV when one hole is injected into the slab, while injections of electrons reduces the MAE. Obviously, the response of the MAE to the charge injection follows the trend predicted by the $E_F$ dependent MAE curve in Fig. \ref{torque2}(a). Therefore, we can expect that the MAE can be even larger than 20 meV per slab when more holes are injected.

\section{Conclusions}
In summary, we studied the electronic and magnetic properties of ultra-thin Fe film on Pb(001), based on first-principles calculations. Huge perpendicular MAE can be produced in $Pb_2/Fe_1/Pb$, with MAE as large as 7.59 meV per effective Fe atom. Electronic structure analysis reveals that small amount of local spin moments are induced on the interfacial Pb atoms. Yet this small spin polarization results in huge MAE due to the large SOC constant of $Pb-6p$ orbitals. Repeating the Pb capping layers and Fe layer by three times [$(Pb_2/Fe_1)_3/Pb$] can lead to even larger MAE of 13.6 meV and it can be tuned by injecting charge. Furthermore, these structures may be fabricated under delicately controlled conditions, so they are of great potential in applications in spintronics devices at high temperature.

\section*{Acknowledgements}

This work is supported by the National Natural Science Foundation of China (11574223), the Natural Science Foundation of Jiangsu Province (BK20150303) and the Jiangsu Specially-Appointed Professor Program of Jiangsu Province.%We also acknowledge the National Supercomputing Center in Shenzhen for providing the computing resources.


\subsection*{REFERENCES}
\begin{thebibliography}{99}
%%%%%%%%%%%%%%%%%%%%%%%%%%%%%%%
%%%%%%%%%%    Introduction    %%%%%%%%%%%%%
%%%%%%%%%%%%%%%%%%%%%%%%%%%%%%%
\bibitem{Science2001} Wolf,S. A.; Awschalom, D. D.; Buhrman, R. A.; Daughton, J. M.; von Moln{\'a}r, S.; Roukes, M. L.; Chtchelkanova, A. Y.; Treger, D. M. Spintronics: A Spin-Based Electronics Vision for the Future. {\it Science} {\bf 2001}, 294, 1488--1495.
\bibitem{RMP2004} \ifmmode \check{Z}\else \v{Z}\fi{}uti\ifmmode \acute{c}\else \'{c}\fi{}, I.; Fabian, J.; Das Sarma, S. Spintronics: Fundamentals and applications. {\it Rev. Mod. Phys.} {\bf 2004}, 76, 323--410.
\bibitem{IEEETM2010} Chen,  E.; Apalkov, D.; Diao, Z.; Driskill-Smith, A.; Druist, D.; Lottis, D.; Nikitin, V.; Tang, X.; Watts, S.; Wang, S.; Wolf,  S. A.; Ghosh, A. W.; Lu, J. W.; Poon, S. J.; Stan, M.; Butler, W. H.; Gupta, S.; Mewes, C. K. A.; Mewes, T.; Visscher, P. B. Advances and Future Prospects of Spin-Transfer Torque Random Access Memory. {\it IEEE Trans. Mag.} {\bf 2010}, 46, 1873--1878.
\bibitem{HeinrichAJ} Loth, S.; Baumannm S.; Lutz, C. P.; Eigler, D. M.; Heinrich, A. J. Bistability in Atomic-Scale Antiferromagnets. {\it Science} {\bf 2012}, 335, 196--199.
\bibitem{LiuLQ} Liu, L. Q.; Pai, C. F.; Li, Y.; Tseng, H. W.; Ralph, D. C.; Buhrman, R. A. Spin-Torque Switching with the Giant Spin Hall Effect of Tantalum. {\it Science} {\bf 2012}, 336, 555--558.
\bibitem{HeinrichBW} Heinrich, B. W.; Braun, L.; Pascual, J. I.; Franke, K. J. Tuning the Magnetic Anisotropy of Single Molecules. {\it Nano Lett.} {\bf 2015}, 15, 4024--4028.
\bibitem{ButlerWH} Butler,W. H.; Zhang, X.-G.; Schulthess, T. C.; MacLaren, J. M. Spin-dependent tunneling conductance of Fe/MgO/Fe sandwiches. {\it Phys. Rev. B} {\bf 2001}, 63, 054416.
\bibitem{HeinrichB-1} Meyerheim, H. L.; Popescu, R.; Kirschner, J.; Jedrecy, N.; Sauvage-Simkin, M.; Heinrich, B.; Pinchaux, R. Geometrical and Compositional Structure at Metal-Oxide Interfaces: MgO on Fe(001).{\it Phys. Rev. Lett.} {\bf 2001}, 87, 076102.
\bibitem{HeinrichB-2} Klaua, M.; Ullmann, D.; Barthel, J.; Wulfhekel, W.; Kirschner, J.; Urban R.; Monchesky, T. L.; Enders,  A.;  Cochran, J. F.; Heinrich, B. Growth, structure, electronic, and magnetic properties of MgO/Fe(001) bilayers and Fe/MgO/Fe(001) trilayers. {\it Phys. Rev. B} {\bf 2001}, 64, 134411.
\bibitem{RMP2017} Dieny, B.; Chshiev, M. Perpendicular magnetic anisotropy at transition metal/oxide interfaces and applications. {\it Rev. Mod. Phys.} {\bf 89}, 025008 (2017).
\bibitem{RE-TM-1} Ohmori, H.; Hatori, T.; Nakagawa, S. Perpendicular magnetic tunnel junction with tunneling magnetoresistance ratio of 64\% using MgO (100) barrier layer prepared at room temperature. {\it J. Appl. Phys.} {\bf 2008}, 103, 07A911.
\bibitem{RE-TM-2} Nakayama, M.; Kai, T.; Shimomura, N.; Amano, M.; Kitagawa, E.; Nagase, T.; Yoshikawa, M.; Kishi, T.; Ikegawa, S.; Yoda, H. Spin transfer switching in TbCoFe/CoFeB/MgO/CoFeB/TbCoFe magnetic tunnel junctions with perpendicular magnetic anisotropy. {\it J. Appl. Phys.} {\bf 2008}, 103, 07A710.
\bibitem{L10-FePt} Shima, T.; Moriguchi, T.; Mitani, S.; Takanashi, K. Low-temperature fabrication of $L1_0$ ordered FePt alloy by alternate monatomic layer deposition. {\it Appl. Phys. Lett.} {\bf 2002}, 80, 288--290.
\bibitem{L10-CoPt} Kim, G.; Sakuraba, Y.; Oogane, M.; Ando, Y.; Miyazaki, T. Tunneling magnetoresistance of magnetic tunnel junctions using perpendicular magnetization $L1_0$-CoPt electrodes. {\it Appl. Phys. Lett.} {\bf 2008}, 92, 172502.
\bibitem{L10-MAE-1} Shick, A. B.; Mryasov, O. N. Coulomb correlations and magnetic anisotropy in ordered $L1_0$ CoPt and FePt alloys. {\it Phys. Rev. B} {\bf 2003}, 67, 172407.
\bibitem{L10-MAE-2} Staunton, J. B.; Ostanin, S.; Razee, S. S. A.; Gyorffy, B. L.; Szunyogh, L.; Ginatempo, B.; Bruno, E. Temperature Dependent Magnetic Anisotropy in Metallic Magnets from an Ab Initio Electronic Structure Theory: $L1_0$-Ordered FePt. {\it Phys. Rev. Lett.} {\bf 2004}, 93, 257204.
\bibitem{L10-MAE-3} Burkert, T.; Eriksson, O.; Simak, S. I.; Ruban, A. V.; Sanyal, B.; Nordstr\"{o}m, L.; Wills, J. M. Magnetic anisotropy of $L1_0$ FePt and Fe$_{1-x}$Mn$_x$Pt. {\it Phys. Rev. B} {\bf 2005}, 71, 134411.
\bibitem{Pt/Fe/Pt-1} Tsujikawa, M; Oda, T. Finite Electric Field Effects in the Large Perpendicular Magnetic Anisotropy Surface Pt/Fe/Pt(001): A First-Principles Study. {\it Phys. Rev. Lett.} {\bf 2009}, 102, 247203.
\bibitem{Pt/Fe/Pt-2} Ruiz-D\'{i}az, P.; Dasa, T. R.; Stepanyuk, V. S. Tuning Magnetic Anisotropy in Metallic Multilayers by Surface Charging: An Ab Initio Study. {\it Phys. Rev. Lett.} {\bf 2013}, 110, 267203.
\bibitem{Fe/W} Qian, X.; H\"{u}bner, W. Ab initio magnetocrystalline anisotropy calculations for Fe/W(110) and Fe/Mo(110).  {\it Phys. Rev. B} {\bf 2001}, 64, 092402.
\bibitem{Fe/Mo} Prokop, J.; Kukunin, A.; Elmers, H. J. Magnetic Anisotropies and Coupling Mechanisms in Fe/Mo(110)  Nanostripes. {\it Phys. Rev. Lett.} {\bf 2005}, 95, 187202.
\bibitem{CoFeB-1}  Kent, A. D. Spintronics: Perpendicular all the way. {\it Nat. Mater.} {\bf 2010}, 9, 699--700.
\bibitem{CoFeB-2} Ikeda, S.; Miura, K.; Yamamoto, H.; Mizunuma, K.; Gan, H. D.; Endo, M.; Kanai, S.; Hayakawa, J.; Matsukura, F.; Ohno, H. A perpendicular-anisotropy CoFeB--MgO magnetic tunnel junction. {\it Nat. Mater.} {\bf 2010}, 9, 721--724.
\bibitem{CoFeB-3} Stewart, D. A. New Type of Magnetic Tunnel Junction Based on Spin Filtering through a Reduced Symmetry Oxide: FeCo|Mg3B2O6/FeCo. {\it Nano Lett.} {\bf 2010}, 10, 263--267.
\bibitem{CoFeB-4} Wang, W. G.; Li, M.; Hageman, S.; Chien, C. L. Electric-field-assisted switching in magnetic tunnel junctions. {\it Nat. Mater.} {\bf 2012}, 11, 64--68.
\bibitem{CoFeB-5} Wang, Z. C.; Saito, M.; McKenna, K. P.; Fukami, S.; Sato, H.; Ikeda, S.; Ohno, H.; Ikuhara, Y. Atomic-Scale Structure and Local Chemistry of CoFeB?MgO Magnetic Tunnel Junctions. {\it Nano Lett.} {\bf 2016}, 16, 1530--1536. %
\bibitem{Freeman1} Hotta, K.; Nakamura, K.; Akiyama, T.; Ito, T.; Oguchi, T.; Freeman, A. J. Atomic-Layer Alignment Tuning for Giant Perpendicular Magnetocrystalline Anisotropy of 3d  Transition-Metal Thin Films. {\it Phys. Rev. Lett.} {\bf 2013},110, 267206.
\bibitem{Ito2015} Nakamuraa, K.; Ikeura, Y.; Akiyama, T.; Ito, T. Giant perpendicular magnetocrystalline anisotropy of 3d transition-metal thin films on MgO. {\it J. Appl. Phys} {\bf 2015}, 117, 17C731.
\bibitem{FCC-Fe} Abrahams, S. C.; Guttman, L.; Kasper, J. S. Neutron Diffraction Determination of Antiferromagnetism in Face-Centered Cubic ($\ensuremath{\gamma}$) Iron. {\it Phys. Rev.} {\bf 1962}, 127, 2052--2055.
\bibitem{SOC-Pb} Wittel, K.; Manne, R. Atomic Spin-Orbit Interaction Parameters from Spectral Data for 19 elements. {\it Theoret. Chim. Acta.} {\bf 1974}, 33, 347--349.
\bibitem{SOC-Pt} Vijayakumar, M.; Gopinathan, M. S. Spin-orbit coupling constants of transition metal atoms and ions in density functional theory. {\it J. Mol. Struct. (Theochem)} {\bf 1996}, 361, 15--19. % 
%%%%%%%%%%%%%%%%%%%%%%%%%%%%%%%
%%%%%%%%%%%%    Method    %%%%%%%%%%%%%
%%%%%%%%%%%%%%%%%%%%%%%%%%%%%%%
\bibitem{VASP1} Kresse, G.; Furthm\"{u}ller, J. Efficiency of ab-initio total energy calculations for metals and semiconductors using a plane-wave basis set. {\it Comput. Mater. Sci.} {\bf 1996}, 6, 15--50.
\bibitem{VASP2} Kresse, G.; Furthm\"{u}ller, J. Efficient iterative schemes for ab initio total-energy calculations using a plane-wave basis set. {\it Phys. Rev. B} {\bf 1996}, 54, 11169--11186.
\bibitem{PAW1} Bl\"{o}chl, P. E. Projector augmented-wave method. {\it Phys. Rev. B} {\bf 1994}, 50, 17953--17979.
\bibitem{PAW2} Kresse, G.; Joubert, D. From ultrasoft pseudopotentials to the projector augmented-wave method. {\it Phys. Rev. B} {\bf 1999}, 59, 1758--1775.
\bibitem{PBE} Perdew, J. P.; Burke, K.; Ernzerhof, M. Generalized Gradient Approximation Made Simple. {\it Phys. Rev. Lett.} {\bf 1996}, 77, 3865--3868.
\bibitem{Torque1} Wang, X. D.; Wu, R. Q.; Wang, D. S.; Freeman, A. J. Torque method for the theoretical determination of magnetocrystalline anisotropy. {\it Phys. Rev. B} {\bf 1996}, 54, 61--64.
\bibitem{Torque2} Hu, J.; Wu, R. Q. Control of the Magnetism and Magnetic Anisotropy of a Single-Molecule Magnet with an Electric Field. {\it Phys. Rev. Lett.} {\bf 2013}, 110, 097202.
\bibitem{Hu-NL-2014} Hu, J.; Wu, R. Q. Giant Magnetic Anisotropy of Transition-Metal Dimers on Defected Graphene. {\it Nano Lett.} {\bf 2014}, 14, 1853--1858.
\bibitem{MAE-Fe} The MAE of Fe thin film does not change notably with the thickness up to 15 MLs, in good agreement with previous calculation \cite{Freeman1}.
%%%%%%%%%%%%%%%%%%%%%%%%%%%%%%%%%
%%%%%%%%%%%%    Discussion    %%%%%%%%%%%%%
%%%%%%%%%%%%%%%%%%%%%%%%%%%%%%%%%
%\bibitem{MAE1} Wang, D. S.; Wu, R. Q.; Freeman, A. J. First-principles theory of surface magnetocrystalline anisotropy and the diatomic-pair model. {\it Phys. Rev. B} {\bf 1994}, 47, 14932.
\bibitem{Yao-bcc-Fe} Yao, Y. G.; Kleinman, L.; MacDonald, A. H.; Sinova, J.; Jungwirth, T.; Wang, D. S.; Wang, E. G.; Niu, Q. First Principles Calculation of Anomalous Hall Conductivity in Ferromagnetic bcc Fe. {\it Phys. Rev. Lett.} {\bf 2004}, 92, 037204.
\bibitem{Tsymbal} Zhang, J.; Lukashev, P. V.; Jaswal, S. S.; Tsymbal, E. Y. Model of orbital populations for voltage-controlled magnetic anisotropy in transition-metal thin films. {\it Phys. Rev. B} {\bf 2017}, 96, 014435.
\bibitem{E-field} Neugebauer, J.; Scheffler, M. Adsorbate-substrate and adsorbate-adsorbate interactions of Na and K adlayers on Al(111). {\it Phys. Rev. B} {\bf 1992}, 46, 16067.
\end{thebibliography}
\end{document}